\documentclass[aps,prl,superscriptaddress,twocolumn,showpacs,amsmath,floatfix]{revtex4-1}

\usepackage{dcolumn}
\usepackage{bm}
\usepackage{ulem}
\usepackage{amsmath,amssymb}
\usepackage{latexsym}
\usepackage{graphicx}
\usepackage{leftidx}
\usepackage{color}

\begin{document}

\title{The origin of hyperferroelectricity in Li$B$O$_3$ ($B$=V, Nb, Ta, Os)}

\author{Pengfei Li}
\affiliation{Key Laboratory of Quantum Information, University of Science and
Technology of China, Hefei, 230026, China}
\affiliation{Synergetic Innovation Center of Quantum Information and Quantum
  Physics, University of Science and Technology of China, Hefei, 230026, China}
\author{Xinguo Ren}
\affiliation{Key Laboratory of Quantum Information, University of Science and
Technology of China, Hefei, 230026, China}
\affiliation{Synergetic Innovation Center of Quantum Information and Quantum
  Physics, University of Science and Technology of China, Hefei, 230026, China}
\author{Guang-Can Guo}
\affiliation{Key Laboratory of Quantum Information, University of Science and
Technology of China, Hefei, 230026, China}
\affiliation{Synergetic Innovation Center of Quantum Information and Quantum
  Physics, University of Science and Technology of China, Hefei, 230026, China}
\author{Lixin He}\email{helx@ustc.edu.cn.}
\affiliation{Key Laboratory of Quantum Information, University of Science and
Technology of China, Hefei, 230026, China}
\affiliation{Synergetic Innovation Center of Quantum Information and Quantum
  Physics, University of Science and Technology of China, Hefei, 230026,
  China}

\date{\today}

\begin{abstract}
{The electronic and structural properties of Li$B$O$_3$ ($B$=V, Nb, Ta, Os) are investigated via first-principles
methods. We show that Li$B$O$_3$ belong to the recently proposed hyperferroelectrics, i.e., they all have
unstable longitudinal optic phonon modes. Especially, the ferroelectric-like instability
in the metal LiOsO$_3$, whose optical dielectric constant goes to infinity, is a limiting case of
hyperferroelectrics.
Via an effective Hamiltonian, we further show that, in contrast to
normal proper ferroelectricity, in which the ferroelectric instability usually comes from
long-range coulomb interactions, the hyperferroelectric instability
is due to the structure instability driven by short-range interactions. This could happen
in systems with large ion size mismatches, which therefore provides a useful guidance in
searching for novel hyperferroelectrics.
}
\end{abstract}

\pacs{77.80.-e, 
77.22.Ej        
}

\maketitle

The switchable polarization of ferroelectrics
made them an important class of
materials for modern device applications.
However, in traditional proper ferroelectrics,
the electric polarization is very sensitive to the domain wall structures and
electric boundary conditions~\cite{Zhong1994}.
This is even more severe in the
case of ferroelectric thin films~\cite{Junquera2003, Sai2009}, where
the depolarization field may easily destroy the polarization states.
Recently in a seminal work~\cite{Garrity2014}, Garrity, Rabe and Vanderbilt (GRV) showed that
a class of recently discovered hexagonal $ABC$ semiconducting ferroelectrics
\cite{Bennett2012}
have very robust polarization properties even when the depolarization field is
unscreened, e.g., can remain polarized down to single atomic layers
when interfaced with normal insulators,
therefore they are given the name of {\it hyperferroelectrics}.
These properties are extremely important for modern
device applications which utilize the ferroelectric
thin films~\cite{Polking2012}.
GRV further showed that the extraordinary behavior of hyperferroelecrics
is because they have an unstable longitudinal optic (LO) mode
besides the transverse optic (TO) mode instability.

It has been proposed that in the hexagonal $ABC$ hyperferroelectrics,
the imaginary LO phonon frequency
is due to the small LO-TO splitting, which further arises from their
small energy gaps -- thus large optical dielectric constants $\epsilon_{\infty}$ --
as well as small Born effective charges~\cite{Garrity2014}.
However, in this Letter,
we show that LiNbO$_3$, LiTaO$_3$ are also hyperferroelectrics, because they
have unstable LO phonon modes as well,
despite that they
have relative small $\epsilon_{\infty}$ and large mode effective charges,
in contrast with the above-mentioned $ABC$ hyperferroelectrics.
This poses an interesting question: is there a more general (fundamental) driving
mechanism for hyperferroelectrics,
besides the small LO-TO splitting scenario disclosed by GRV for hexagonal $ABC$ ferroelectrics?
We will answer the question in this work using the Li$B$O$_3$-type materials.

\begin{figure}
\centering
\includegraphics[width=2.0in]{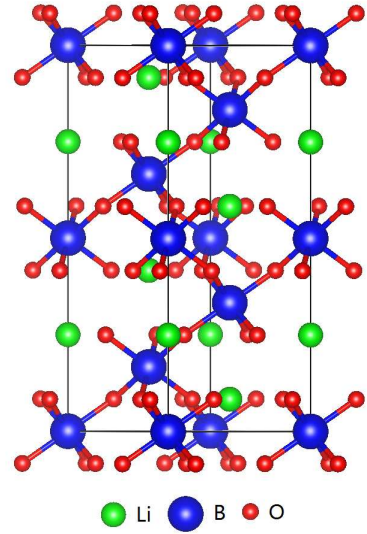}
\caption {(Color online)
The hypothetical paraelectric (PE) structure
of Li$B$O$_3$ resulting
from an average of the disordered structure above $T_c$.
}
\label{fig:structure}
\end{figure}

LiNbO$_3$ and LiTaO$_3$ are two important ferroelectrics
which have been investigated intensively in the past years
\cite{Penna1976,Okamoto1985,Zhang1986,Catchen1991,Cheng1993,Inbar1996}.
The ferroelectric transition of these materials is believed
to be of order-disorder character.
Shown in Fig.~\ref{fig:structure} is the hypothetical paraelectric (PE) structure
of Li$B$O$_3$ ($B$=Nb, Ta), resulting
from an average of the disordered structure above $T_c$.
The paraelectric (PE) structure belongs to the $R\bar{3}c$ space group, whereas
the ferroelectric (FE) structure is rhombohedral, and belongs to the space group $R3c$.
In the FE phase, the Li, O and $B$ ions distort from their central symmetric positions,
and induce the electric polarization along the trigonal axis.
Inbar and Cohen studied the electronic
and structural properties of LiNbO$_3$ and LiTaO$_3$~\cite{Inbar1996}.
They found large hybridization between the transition-metal $B$
atoms and the oxygen 2$p$ states, similar to perovskite ferroelectrics.
It has thus been suggested that the ferroelectricilty in
LiNbO$_3$ and LiTaO$_3$ is due to long-range Coulomb interactions.
Interestingly, very recently it has been found that LiOsO$_3$, even though being
a metal, also has ferroelectric-like structural transitions~\cite{Shi2013}. It is
very puzzling where the ferroelectric-like structure of LiOsO$_3$ comes from,
since the long-range Coulomb long-range interactions
should be screened in the metallic states. The mechanism for the
FE-like structural transition in
metal is still under debate~\cite{Xiang2014,Liu2015}.

In this Letter, we investigate the ferroelectric properties of
Li$B$O$_3$-type compounds, where $B$=V, Nb, Ta and Os.
We show that Li$B$O$_3$ are hyperferroelectrics, and
there are two co-existing and yet distinct
ferroelectric mechanisms in Li$B$O$_3$,
namely the long-range Coulomb interactions
due to $B$ ions, and short-range structural instability due to Li ions.
Especially we show that the instability of Li ions is responsible for
the hyperferroelectric behavior of Li$B$O$_3$.
The FE-like structural transition in metallic LiOsO$_3$~\cite{Shi2013}
is nothing special, but has the same mechanism of other Li$B$O$_3$ compounds.
In this sense, LiOsO$_3$ can be viewed as a special hyperferroelectrics
in the limit of $\epsilon_{\infty}$$\rightarrow$$\infty$.
Via an effective Hamiltonian model, we further clarify that the microscopic origin of
hyperferroelectrics is from the instability driven by short-range interactions.
These results provide a strong guidance in searching for novel hyperferroelectrics.

\begin{table}
\caption{Calculated band gaps, optical dielectric constants and atomic
Born effective charges of Li$B$O$_3$.}
\centering
\begin{tabular}{lccccc}
\hline
\hline
& gap (eV) &$\epsilon_{\infty}$  & $Z_c^*$(Li) & $Z_c^*$(B) & $Z_c^*$(O)\\
\hline
LiVO$_3$  &0.4 & 18.8 & 1.13 &13.36  & -4.83 \\
LiNbO$_3$ &2.2 & 7.3 & 1.09 & 9.37 & -3.49 \\
LiTaO$_3$ &3.0 & 5.8 & 1.10 & 8.36 & -3.15 \\
LiOsO$_3$ & 0  &$\infty$ & \\
\hline \hline
\label{tab:epsilon}
\end{tabular}
\end{table}

The ferroelectric phase transitions can be understood by the lattice
dynamics of their high-symmetry phase.
For ferroelectrics, the high-symmetry phase has at least one unstable
TO mode.
The frequencies of TO can be calculated using first-principles methods
in bulk materials in the absence of macroscopic electric field ($E$=0)~\cite{Garrity2014}.
If the depolarization field is unscreened, corresponding to the case of
electric displacement $D$=0, the structure instability is determined
by the LO modes, which can be
obtained by adding to the
dynamic matrix a non-analytic
long-range Coulomb term (known as the LO-TO splitting)~\cite{Pigk1970,Zhong1994}
that schematically takes the form $4\pi {Z^*}^2/\Omega\epsilon_{\infty}$,
where $Z^*$ is the Born effective charge
and $\Omega$ is the volume of unit cell.
In normal ferroelectrics, such as PbTiO$_3$, BaTiO$_3$, etc.,
due to their large Born effective charges and relatively small
$\epsilon_{\infty}$, the LO-TO splittings are huge, such that
all LO modes are stable~\cite{Zhong1994}.
Therefore they lose ferroelectricity
if the depolarization field is not well screened.
In constrast, in the $ABC$ hexagonal hyperferroelectrics (e.g., LiZnAs)
as discussed by GRV~\cite{Garrity2014}, the LO-TO splittings are small, such that
even the LO modes can become unstable.
Consequently, the polarization in these materials is very robust against
the depolarization field.

Ferroelectric materials with small LO-TO splittings are the most
obvious candidates for hyperferroelectrics.
Therefore, it is a natural attempt to look for the hyperferroelectrics
in materials with (i) small band gap, or equivalently
large electronic dielectric
constant $\epsilon_{\infty}$; (ii) small mode effective charges.
Indeed, the hyperferroelectrics found by GRV all satisfy
these conditions~\cite{Garrity2014}.
However, as demonstrated below, LiNbO$_3$  and LiTaO$_3$
are also hyperferroelectrics, i.e., having unstable LO phonon modes,
despite that they have large band gap, relatively small optical dielectric constants,
and large effective mode effective charges.

The electronic and structural properties of
LiBO$_3$ are calculated using density functional
theory within local density approximation (LDA),
implemented in the Vienna
ab initio simulations package (VASP)~\cite{kresse93,kresse96}.
The projector
augmented-wave (PAW) pseudopotentials~\cite{blochl94}
with a 500 eV plane-wave cutoff are used.
The Brillouin zone is sampled with a 8$\times$8$\times$8
Monkhorst-Pack k-point grid converges the
results very well.
We relax the structure until
the remaining forces are less than 1 meV/\AA.
Phonon frequencie
are calculated using a finite difference method
as implemented
in Phonopy package~\cite{phonopy}.
The Born effective charges and the optical dielectric constants  are
calculated using density functional perturbation theory (DFPT)
\cite{Gajdo2006}.
The above properties of
LiBO$_3$ are also calculated using other functionals,
including GGA and LDA+U.
Although the exact numbers of the results may vary, the main conclusions
are remain unchanged.

The calculated band gaps (via DFT-LDA) of Li$B$O$_3$
are listed in Table~I
for their paraelectric states.
We see that the LDA calculated
band gaps of LiNbO$_3$ and LiTaO$_3$ are 2.2 and 3.0 eV respectively
(which suprprisingly are not much underestimated compared to the
experimental values~\cite{Thierfelder2010,Polshettivar2013}).
The band gaps are comparable to those of perovskite ferroelectrics,
much larger than the band gaps of $ABC$ hexagonal hyperferroelectrics,
which are around 0.5 -
1 eV~\cite{Garrity2014}.
LiVO$_3$ has relative small LDA band gap, which is approximately 0.4 eV.
LiOsO3 is a metal~\cite{Shi2013}.
The calculated optical dielectric constant $\epsilon^{zz}_{\infty}$
along the (polar) $z$ axis
and Born effective charges are also listed in Table~I.
We see that LiNbO$_3$ and LiTaO$_3$ have $\epsilon^{zz}_{\infty}$ approximately 5 -7, similar
to those of peroskite ferroelectrics, e.g., PbTiO$_3$, BaTiO$_3$. The dielectric constant of LiVO$_3$
is somehow larger, approximately 19,
consistent with the fact that LiVO$_3$ has smaller band gap. LiOsO$_3$ is
a metal, and therefore its $\epsilon^{zz}_{\infty}$ diverges.
We then calculate the atomic Born effective charges for LiVO$_3$, LiNbO$_3$ and LiTaO$_3$.
The effective charge of Li ions is approximately
1.0, suggesting that Li is totally ionized.
The effective charges of V, Nb, and Ta ions
are approximately 13, 9, and 8 respectively, which are
anomalously larger than their valence electron charges,
similar to those of peroskite ferroelectrics. 
It is usually believed that the anomalous effective charges introduce
large long-range Coulomb interactions,
leading to spontaneous electric polarization in ferroelectrics
\cite{Inbar1996,Zhong1995}.

\begin{table}
\caption{Calculated phonon frequencies
of the softest TO modes ($\omega_{TO}$), LO modes
($\omega_{LO}$) and the phonon modes due to pure short-range
interactions [$\omega_{s}$, calculated using Eq.~(\ref{Eq:Ds})]. Also shown are
 the mode effective charges  $Z^*_{TO}$ of the TO modes.
and the electric polarization under $E$=0 and $D$=0. The
values for PbTiO$_3$, BaTiO$_3$, NaNbO$_3$, and KNbO$_3$
are obtained under their cubic structures.}
\centering
\begin{tabular}{lcccccccc}
\hline
\hline
& $\omega_{TO}$ & $\omega_{LO}$ & $\omega_s$ & $Z^*_{TO}$ &$P_{E=0}$ &
$P_{D=0}$ \\
  & (cm$^{-1}$) & (cm$^{-1}$) &  (cm$^{-1}$) &  & (C/m$^2$) & (C/m$^2$) \\
\hline
LiVO$_3$  & 409$i$  &  160$i$ &160$i$   & 20.1 & 1.79 & 0.29 \\
LiNbO$_3$ & 208$i$ & 104$i$  & 125$i$    & 8.9  & 1.00 & 0.08\\
LiTaO$_3$ &188$i$  &  77$i$  & 110$i$    & 5.8  & 0.72 & 0.05\\
LiOsO$_3$ &183$i$  &183$i$   &  183$i$   &  -   &  -   & -\\
PbTiO$_3$ &119$i$  &105   &95     &7.5      &0.57     &0\\
BaTiO$_3$ &91$i$ &181 &180 &10.1 &0.12 &0 \\
NaNbO$_3$ &167$i$ &81 &73 &8.7 &0.49 &0   \\
KNbO$_3$ &143$i$ &172 &171 &10.8 &0.29 &0 \\
\hline
\hline
\label{tab:summary}
\end{tabular}
\end{table}

To study the structure instabilities in Li$B$O$_3$,
we calculate the phonons of Li$B$O$_3$
at $\Gamma$ point in the PE phases using a 10-atom unit cell.
We focus on the $A_{2u}$ modes which are associated with the
ferroelectric structural transitions.
For these modes, the phonon frequencies for
both the TO mode and LO modes are calculated.
The summary of the results are given in Table II. As a comparison,
we also present the results for some normal peroskite ferroelectrics,
including PbTiO$_3$, BaTiO$_3$, NaNbO$_3$, and KNbO$_3$. We note that
the results for the peroskites are all obtained at their cubic structure.
As listed in Table II, LiVO$_3$, LiNbO$_3$, LiTaO$_3$, and LiOsO$_3$
all have very strong instable TO modes.
This is consistent with the proposal that LiTaO$_3$ and LiNbO$_3$
are order-disorder ferroelectrics~\cite{Penna1976,
Okamoto1985,Zhang1986,Catchen1991,Cheng1993},
in which the centrosymmetric structures have much higher
energies than the distorted structures.
The calculated mode effective charges for LiVO$_3$, LiNbO$_3$, and LiTaO$_3$
are approximately 20, 9, and 6, respectively.
The Born effective charges are ill-defined for LiOsO$_3$, which is a metal.
We also present the electric polarizations for their FE phase, which are comparable
to those of perovskite ferroelectrics.

\begin{figure}
\centering
\includegraphics[width=2.8in]{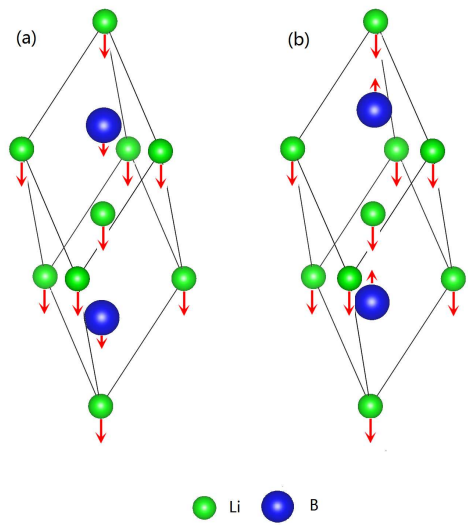}
\caption {(Color online) The schematic phonon patterns
showing the atomic displacements of Li and $B$
atoms for the soft (a) TO mode, and (b) LO mode.
For clarity, we neglect the oxygen atoms.}
\label{fig:phonon}
\end{figure}

The LO phonon frequencies are calculated by diagonalizing
the resultant matrix obtained by
adding the non-analytic terms to the dynamic matrix, i.e.,
\begin{equation}
D^{\rm LO}_{ij}(0)=D^{\rm TO}_{ij}(0) +
{4\pi \over \Omega} {Z_i^*Z_j^* \over \epsilon_{\infty}} \, ,
\label{eq:LO-mode}
\end{equation}
where $Z_i^*$, $Z_j^*$ are the atomic effective charges.
The results are also given in Table~\ref{tab:summary}.
Remarkably, all calculated Li$B$O$_3$
compounds have soft LO modes, indicating that they are hyperferroelectrics,
similar to the $ABC$ hexagonal ferroelectrics but in contrast to the peroskite ones.
For LiOsO$_3$, the TO modes and LO modes have the same frequencies, because its
$1/\epsilon_{\infty}$=0.
These results are quite surprising, given
that the dielectric constants $\epsilon_{\infty}$
of LiNbO$_3$ and LiTaO$_3$ are relative small, and
the mode effective charges are quite large, similar
to those of the traditional perovskite
ferroelectrics, such as PbTiO$_3$, BaTiO$_3$ etc.
One may expect that the LO-TO splitting
$4\pi {Z^*}^2/\Omega\epsilon_{\infty}$ would stabilize all LO
modes.
To understand the origin of the soft LO modes,
we analyze the eigenvectors of the soft $A_{2u}$ modes of Li$B$O$_3$,
for both TO modes and LO modes.
The atomic displacements of the soft TO
and LO phonons are shown in Fig.~\ref{fig:phonon}(a) and~\ref{fig:phonon}(b),
respectively.
In TO modes, Li ions and $B$ ($B$=V, Nb, Ta)
ions move in the same direction,
whereas the O ions (not shown) move along the opposite direction.
The Li ions have the largest displacement, where
$B$ and O ions also have significant contributions.
For LO modes, the displacements of O ions along the $c$
axis are somehow suppressed. Surprisingly, the displacements of $B$
ions reverse from those of the TO mode, i.e.,
opposite to the polarization direction!
These results show that the phonon eigenvectors
are very sensitive to the electric boundary conditions, and are very different
for TO modes and LO modes.
Therefore, adding the simple correction
term $4\pi {Z^*}^2/\Omega\epsilon_{\infty}$ to the TO modes
would significantly
overestimate the LO-TO splitting, and falsely stablizes all LO modes.
One has to make the non-analytical corrections
to the dynamic matrices themselves.
As we see from Eq.~\ref{eq:LO-mode}
the corrections to the V, Nb, Ta ions are very large due to their large effective charges;
whereas the corrections for Li ions are small,
because $Z^*({\rm Li}) \approx$1, is small.
Therefore, the LO modes of Li$B$O$_3$ can remain soft by altering their
mode patterns.

From the above analysis, one can see
there are two co-existing
ferroelectric mechanisms in Li$B$O$_3$.
One is the long-range Coulomb interactions due to
large Born effective charges arising from the $B$ ions, which are
very sensitive to the electric boundary conditions, just like normal
proper ferroelectrics;
the second is short-range instability due to the large size mismatch between the Li
and $B$ ions~\cite{Xiang2014}, which is
robust against the electric boundary conditions.
Especially for metallic LiOsO$_3$, the FE-like
structural transition is induced by the short-range instability
of Li alone, because the long-range
Coulomb interactions is screened.

We also calculated the electric polarization under the condition $D$=0 by using the model of Ref.~\cite{Garrity2014},
and the results are listed in Table~\ref{tab:summary}. As expected, for normal ferroelectrics PbTiO$_3$, BaTiO$_3$,
NaNbO$_3$, and KNbO$_3$, the spontaneous polarizations are all zero under $D$=0. In contrast,
for LiVO$_3$, LiNbO$_3$, and LiTaO$_3$, the spontaneous polarizations under $D$=0 are about one tenth of
those under $E$=0, but still significant for applications.
These results further confirm that they are hyperferroelectrics.


We have shown that the large electronic dielectric constant and small
effective charges may not be the necessary condition for the hyperferroelectrics.
This raises an interesting question: what are responsible for it?
To answer this question, we start from a simplified effective Hamiltonian
for ferroelectrics following Ref. \cite{Zhong1995},
\begin{equation}
E(\{{\bf u}_i\})=E^{\rm dipol}(\{{\bf u}_i\})
+ E^{\rm self}(\{{\bf u}_i\}) + E^{\rm short}(\{{\bf u}_i\})\, ,
\end{equation}
where ${\bf u}_i$ are the local normal modes at $i$-th cell.
$E^{\rm dipol}$ represents the long-range dipole-dipole interaction,
whereas $E^{\rm self}$, $E^{\rm short}$ are the energies of isolated local modes,
and the short-range interactions between the local modes respectively.
For the simplicity of discussion, we neglect the elastic energies,
and their coupling to the local modes. Without losing generality,
we further assume that the crystal has simple cubic structure.

First, let's look at the dipole-dipole interactions,
%
\begin{equation}
E^{\rm dipol}(\{{\bf u}_i\})={{Z^*}^2 \over \epsilon_{\infty}} \sum_{i<j}
{{\bf u}_i \cdot {\bf u}_j - 3(\hat{{\bf R}}_{ij} \cdot {\bf u}_i) (\hat{{\bf R}}_{ij} \cdot {\bf u}_j) \over
 {R}^3_{ij}} \, ,
\label{eq:dipole}
\end{equation}
where $\epsilon_{\infty}$ is the optical dielectric constant of the material.
$R_{ij}$=$|{\bf R}_{ij}|$ is the distance between the two local modes, where
${\bf R}_{ij}$=${\bf R}_{i}-{\bf R}_{j}$
and $\hat{{\bf R}}_{ij}$=${\bf R}_{ij}/R_{ij}$.
Direct evaluation of Eq.~\ref{eq:dipole} in real space converges very slowly.
Equation~\ref{eq:dipole} can be evaluated using Ewald summation techniques.
For simple cubic structure of infinite lattice size, the summation have been obtained
in Ref.\cite{Cohen1955}.
It turns out that $E^{\rm dipol}$ is non-analytic when ${\bf q} \rightarrow$0, ,
\begin{equation}
E^{\rm dipol}(\{{\bf u}\})=-{{2\pi \over 3\Omega}} {{Z^*}^2 \over \epsilon_{\infty}}
\left(1-3{q_z^2 \over |{\bf q}|^2} \right)u^2
\end{equation}
where $u=|{\bf u}|$ and $\Omega$ is the unit cell volume. Here, we assume that the phonon displacements
are along the $z$ axis.
The short-range interactions can be obtained by setting $Z^*\rightarrow$0,
or $\epsilon_{\infty} \rightarrow \infty$.
The self-energy and the energy due to short-range interactions can thus be written in the
following form as
${\bf q} \rightarrow$0,
\begin{equation}
E^{\rm self}(\{{\bf u}_i\})+E^{\rm short}(\{{\bf u}_i\})=E_0
+{1\over 2}\omega^2_s u^2 + {1\over4}\kappa_4 u^4\, ,
\end{equation}
where $\omega^2_s$=$\kappa_2 + {1 \over 2}\sum_{ij} J_{ij}$,
$\kappa_2$ is
the on-site enenrgy contribution, and $J_{ij}$ are the coupling constants between local modes ${\bf u}_i$
and ${\bf u}_j$.
Therefore, the phonon frequency of the TO mode can be calculated as $q_x$, $q_y \rightarrow$0,
\begin{equation}
\omega^2_{TO}=\omega^2_s-{{4\pi \over 3\Omega}}{{Z^*}^2 \over \epsilon_{\infty}}  \, ,
\label{eq:TO}
\end{equation}
and the phonon frequencies of LO modes can be obtained as,
\begin{equation}
\omega^2_{LO}=\omega^2_s + {{8\pi \over 3\Omega}}{{Z^*}^2 \over \epsilon_{\infty}} \, ,
\label{eq:LO}
\end{equation}
i.e., $\omega^2_{LO}$=$\omega^2_{TO}+{{4\pi \over\Omega}}{{Z^*}^2 \over \epsilon_{\infty}}$.
Assuming that the eigenvectors of LO modes do not change much from that of TO modes,
we can estimate $\omega^2_s$=$(2\omega^2_{TO}+\omega^2_{LO})/3$.
More generally, $\omega^2_s$ can be obtained by solving the following
dynamic matrix,
\begin{equation}
D^{s}_{ij}(0)=D^{\rm TO}_{ij}(0) + {4\pi \over 3\Omega} {Z_i^*Z_j^* \over \epsilon_{\infty}} \,.
\label{Eq:Ds}
\end{equation}

We list the softest $\omega_s$ for typical
perovskite ferroelctrics as well as LiBO$_3$, in Table \ref{tab:summary}.
As we can see from the Table,
in traditional perovskite ferroelectrics, such as
PbTiO$_3$, BaTiO$_3$ etc., $\omega_s$ are all stable,
meaning that the short-range interactions favorite
the high symmetry non-polar structures.
However, because of the large Born effective charges and small optical
dielectric constants $\epsilon_{\infty}$
in these materials,
the long-range Coulomb
interactions (the second term in the right hand of Eq.~\ref{eq:TO})
overcome the short-range repulsive interactions $\omega^2_s$,
and the TO phonon mode frequencies become soft. In these materials, LO modes $\omega^2_{LO}$
are all positive because  $\omega^2_s$ are already positive.
These results are consistent with those of Ref.~\onlinecite{Zhong1994},
and the early point of view
that the long-range Coulomb interactions are the
driven forces for the ferroelectric states \cite{Zhong1995}.

However, for Li$B$O$_3$, because the LO modes are soft (i.e., $\omega^2_{LO}<$0),
it easy to see from Eq.~\ref{eq:LO} that $\omega_s^2$ must also be negative.
This suggests that the short-range interactions already favor the symmetry-broken polarized state
in these materials.
We therefore obtain one of the
most important results of this paper: hyperferroelectrics are
a class of ferroelectrics, where the short-range
interactions already favorite the symmetry broken polar states.
This is a general feature of hyperferroelectrics, or more precisely, a necessary
condition for hyperferroelectrics.
This could happen in materials, e.g., Li$B$O$_3$, where
the ions have large size mismatches. Since the hyperferroelectricity comes from
short-range local interactions, hyperferroelectrics are not sensitive to the
electric boundary conditions.
Especially, LiOsO$_3$ can be viewed as a special hyperferroelectrics,
in which $\epsilon_{\infty} \rightarrow \infty$. More generally, any FE instability
in a metal is a limiting case of hyperferroelectric. It is interesting to see if more
such metals can be found in searching for novel hyperferroelectrics.

To conclude, we have
shown that Li$B$O$_3$ ($B$=V, Nb, Ta, Os)
belong to the recently proposed hyperferroelectrics, despite that some of them (LiNbO$_3$ and LiTaO$_3$)
have large band gaps and Born effective charges.
By resorting to an effective Hamiltonian model,
we clarify that
the origin of the hyperferroelectrics is due to the structural
instability driven by the short-range interactions. At least one route to
find hyperferroelectrics is to search in materials with
large ion size mismatches.
This work therefore provides a useful guidance
in searching for novel hyperferroelectrics.

LH acknowledges valuable discussions with Prof. David Vanderbilt.
The work is supported by the Chinese National Science Foundation Grant number 11374275.


\end{document}